\documentclass{an}
\usepackage{graphicx}
\usepackage{times}
\usepackage{epsf}
\usepackage{psfig}
\Volume{00}              
\Year{0000}              
\Month{00}               
\Pagespan{000}{000}      

\begin{document}
\lhead[\thepage]{A.N. S.F.S\'anchez \& C.R.Benn: Impact of astronomical research from different countries}
\rhead[Astron. Nachr./AN~{\bf XXX} (200X) X]{\thepage}
\headnote{Astron. Nachr./AN {\bf 32X} (200X) X, XXX--XXX}

\title{Impact of astronomical research from different countries}

\author{S.F. S\'anchez\inst{1,2} C.R. Benn\inst{1}}
\institute{Isaac Newton Group, Apt. 321, 38700 Santa Cruz de La Palma, Spain \and Astrophysikalisches Institut Potsdam, And der Sternwarte 16, 14482  Potsdam, Germany}
\date{Received {Nov 11, 2003}; 
accepted {Jan 6, 2004}} 

\abstract{
The impact of astronomical research carried out by different countries
has been compared by analysing the 1000 most-cited astronomy papers
published 1991-8 (125 from each year).  61\% of the citations are to
papers with first authors at institutions in the USA, 11\% in the UK,
5\% in Germany, 4\% in Canada, 3\% in Italy and 3\% in France.  17\%
are to papers with first authors in ESO countries.  The number of
citations is approximately proportional to the number of IAU members
in a given country.  The number of citations per IAU astronomer is
highest in the USA, Switzerland and the UK.  Within continental
Europe, the number of citations per IAU astronomer varies little from
country to country, but is slightly higher in the north than in the
south.  The sample of 1000 papers maps regional subject preferences.
62\% of the extragalactic papers in the sample were published from the
USA, 15\% from the UK, 23\% from other countries (mainly in
continental Europe).  62\% of the papers on stars were also published
from the USA, but the fractions from the UK and from other countries
are 2\% and 36\% respectively.
 }
\correspondence{ssanchez@aip.de}

\maketitle

\section{Introduction}

Nearly all developed countries support vigorous programmes of research
in astronomy, but there have been few comparisons of the relative
scientific impact of astronomical research in different countries.
Research environments and strategies differ from country to country,
so a detailed comparison may shed light on which policies are more
successful.

A list of the 125 most-cited astronomy/space papers for each year 1991-8, 1000
papers in all, was purchased from the Institute for Scientific Information
(ISI) in Philadelphia.  In Benn \& Sanchez (2001, hereafter Paper I), we used
these data to compare the scientific impact of different telescopes.  Here we
use the same dataset to compare the scientific impact of research from
different countries.  See Paper I for details of the sample and analysis, and
also for a discussion of the various biases which affect citation anlayses.
Biases against both publication and citation need to be considered when
comparing citation counts for different countries.  Language bias operates at
a number of levels.  (1) A requirement to publish in English favours native
speakers of the language.  (2) English-speaking scientists tend not to read or
cite papers written in other languages (e.g. Rees 1997, Nature 2002b).  (3)
Citation databases provide uneven coverage of foreign-language journals (e.g.
Ziman 2001, Moed 2002).  A separate bias arises from the tendency of each
community to over-cite its own results, e.g. through preferentially reading
and citing national journals (e.g. Durrani 2000 found that papers from large
countries receive more citations than papers from small countries).  These
biases are likely to favour over-citation of papers from the large north
American and UK astronomy communities.

\section{Citation impact by country}


\begin{table*}
\begin{center}
\caption{Citation impact by country, for all countries with at least one 
paper in the top-cited 1000 published 1991-8}
\begin{tabular}{llrrrrrrrrr}
\hline
Country&ISO&$\Sigma$Papers&$\Sigma$Citns\%&All astro&IAU&Citns\%&Science&GDP&Popn.&IAU/\\
       &code&top-1000&top-1000&papers\% &mem&/100 IAU&\%&$\$$10$^9$&10$^6$&
       10$^6$ popn.\\
(1) & (2) & (3) & (4) & (5) & (6) & (7) & (8) & (9)& (10) & (11) \\
\hline
Australia &AU&   18&  1.6&    3.3& 191&0.8&  2.2&   393&   18.8&  10.2\\
Austria   &AT&   1&   0.1&       &  31&   &  0.7&   184&    8.1&   3.8\\
Belgium   &BE&   2&   0.2&    1.1&  88&   &  1.1&   236&   10.2&   8.6\\
Brazil    &BR&   2&   0.1&       & 109&   &  0.6&  1035&  171.1&   0.6\\
Canada    &CA&  40&   3.8&       & 199&1.9&  4.3&   688&   31.0&   6.4\\
Chile     &CL&   5&   0.4&       &  46&   &  0.2&   184&   15.0&   3.1\\
Denmark   &DK&   6&   0.7&       &  52&1.3&  1.0&   124&    5.4&   9.6\\
Estonia   &EE&   1&   0.1&       &  22&   &  0.0&     7&    1.4&  15.7\\
Finland   &FI&   1&   0.1&       &  37&   &  0.8&   103&    5.2&   7.1\\
France    &FR&  35&   2.9&    8.7& 609&0.5&  5.7&  1320&   59.0&  10.3\\
Germany   &DE&  56&   4.9&   11.7& 488&1.0&  7.2&  1813&   82.1&   5.9\\
Israel    &IL&   3&   0.3&    1.1&  45&   &  1.1&   101&    5.8&   7.8\\
Italy     &IT&  31&   3.2&    7.2& 409&0.8&  3.4&  1181&   56.7&   7.2\\
Japan     &JP&  23&   2.4&    5.1& 448&0.5&  8.2&  2903&  126.2&   3.5\\
Netherlands 
          &NL&   23&  2.1&    3.9& 167&1.3&  2.3&   348&   15.8&  10.5\\
Poland    &PL&   4&   0.4&       & 117&   &  0.9&   263&   38.6&   3.0\\
Russia    &RU&   2&   0.1&       & 344&   &  4.1&   593&  146.4&   2.3\\
S.Africa  &ZA&   3&  0.4&        &  46&   &  0.4&   290&   43.4&   1.1\\
S.Korea   &KR&   1&  0.1&     0.4&  51&   &  0.5&   585&   46.9&   1.1\\
Spain     &ES&   7&   0.8&    4.4& 204&0.4&  2.0&   645&   39.2&   5.2\\
Sweden    &SE&   4&   0.6&       &  95&   &  1.8&   175&    8.9&  10.7\\
Switzerland
          &CH&  14&   1.9&       &  70&2.7&  1.6&   191&    7.3&   9.6\\
UK        &UK& 101&  10.7&   10.3& 535&2.0&  7.9&  1252&   59.1&   9.1\\
Ukraine   &UA&   2&   0.2&       & 119&   &  0.6&   108&   49.8&   2.4\\
USA       &US& 599&  60.6&   42.9&2235&2.7& 30.8&  8511&  272.6&   8.2\\
Venezuela &VE&   2&   0.3&       &   9&   &  0.1&   194&   23.2&   0.4\\
          &&& &&& &&& \\
ESO       &  & 171&  16.5&   32.6&1978&0.8& 24.1&  5388&  245.4&   8.1\\
\hline
\end{tabular}
\end{center}
Columns 1 -- 2 give the country name and ISO code (the latter used as
a plot symbol in Figs. 1 and 2).  Column 3 -- 4 give the number of the
top-cited 1000 articles with first author hosted by the given country,
and the citation fraction (column 4 sums to 100\%).  Column 5 gives
the percentage of {\it all} astronomical articles published (from the
ISI web pages).  Columns 6 -- 7 give the number of IAU members (from
the IAU bulletin of June 1998) and, for countries with more than 5
papers (column 3), the citation fraction (column 4) per 100 IAU
members.  Columns 8 -- 11 give the all-science citations fraction
(Gibbs 1995, similar numbers are presented by May 1997), annual gross
domestic product (GDP), total population (CIA, 1999), and number of
IAU members per million population.  The last line of the table is a
sum over ESO countries (Belgium, Denmark, France, Germany, Italy, the
Netherlands, Sweden and Switzerland, in 1999).  14 of the 1000 papers
could not be classified by country (mostly because they were published
in solar-physics journals to which we did not have access).
\end{table*}

\begin{figure}
\begin{center}
\centering
\psfig{file=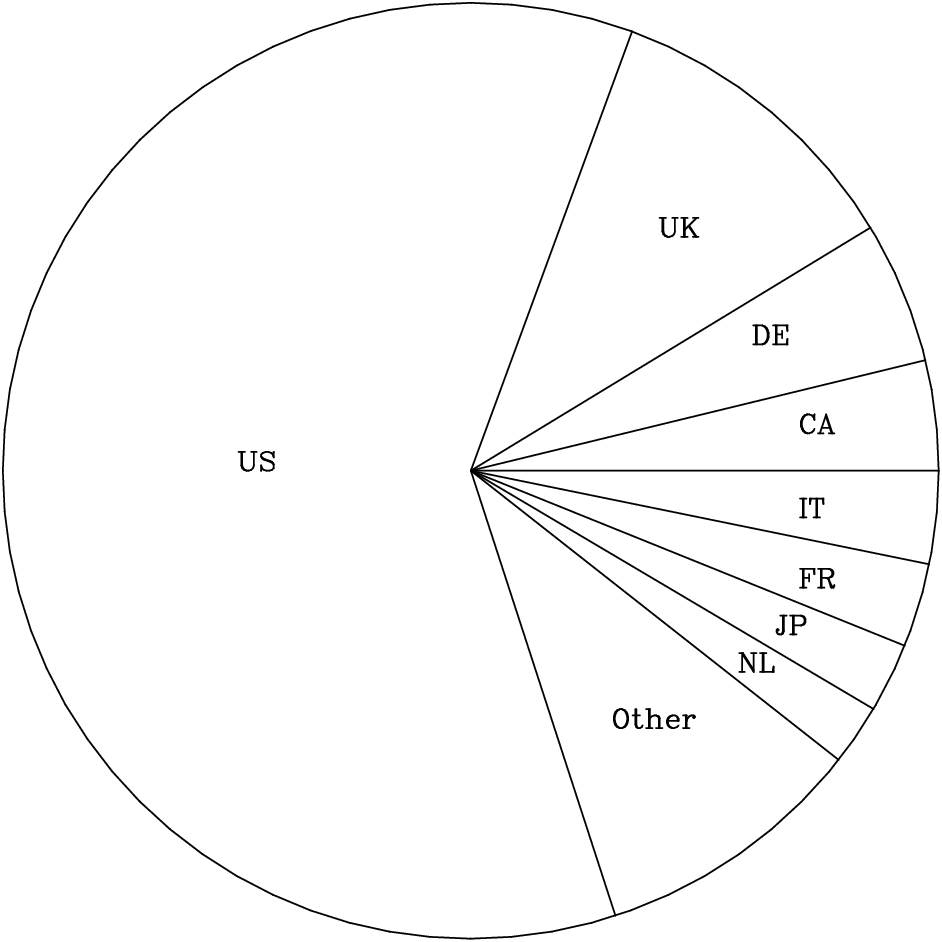,width=7.5cm,height=7.5cm,angle=270}
\end{center}
\caption{
Citation fractions for papers published 1991-8 from each country.
The countries are  represented by their 2-character ISO codes, as given in
Table 1.
}  
\end{figure}

\begin{figure*}
\centering
\psfig{file=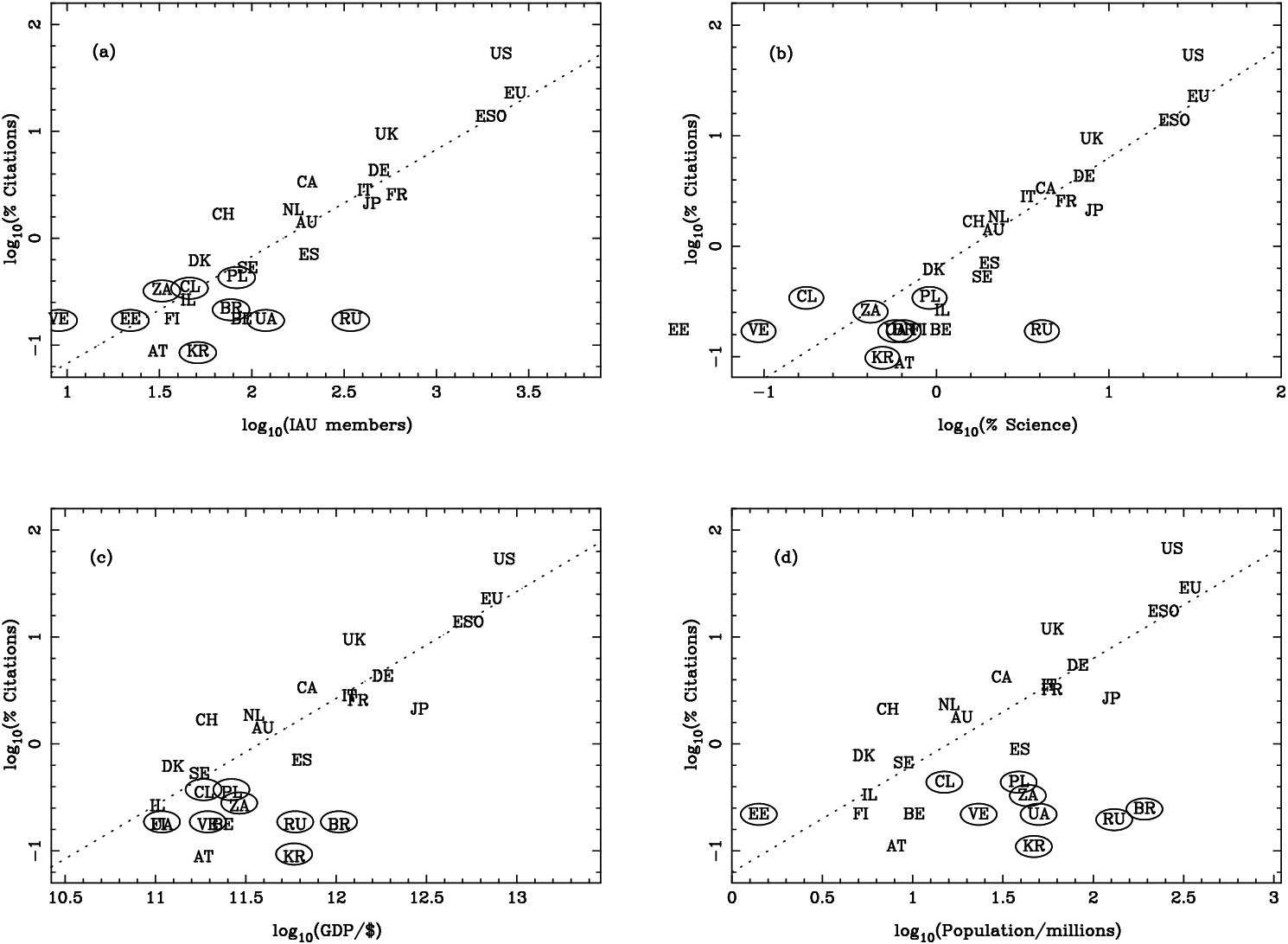,width=17cm,height=13.5cm,angle=270}
\caption{
Citation fraction 1991-8 for each country, compared with 
(a) IAU membership,
(b) world share of all science citations,
(c) gross domestic product (GDP) in $\$$,
(d) country population.
The dotted lines have slope 1 and their vertical positions are arbitrary.
The point `ESO' is a sum over the (then)
8 ESO member countries: BE, DK, FR, DE, IT, NL, SE, CH.
The point `EU' is a sum over the European Union countries included
in Table 1: BE, DK, FR, DE, IT, NL, PT, ES, SE and UK.
Circles indicate countries with GDP per capita
$<$ $\$$15000 per year.
}
\end{figure*}

Each of the 1000 top-cited papers was credited to the country of the
institution hosting the first author.  The number of papers generated
by each country, and the corresponding fractions of citations are
given in Table 1.  The breakdown of citation fractions by country is
shown in Fig. 1.  In Fig. 2, we compare the citation fractions for
each country with four measures which are likely to correlate with the
resources invested by that country in astronomical research: the
number of IAU members; the country's share of all-science citations;
gross domestic product (GDP); and total population (CIA, 1999).  The
number of citations is approximately proportional to the number of IAU
members in each country, and to the country's all-science citation
share, over 2 orders of magnitude (Figs. 2a, b).  The correlations
with GDP and total population (Figs. 2c, d) are weaker (i.e., larger
dispersion), as one would expect, given that countries spend different
fractions of their wealth on research.  In particular, research in
countries with GDP per capita $<$ $\$$15000 (circled on Fig. 2) yields
fewer citations per unit GDP, i.e. these countries probably invest a
smaller fraction of GDP in astronomical research.

The USA dominates, receiving 61\% of the citations to the top 1000
papers, considerably higher than its all-science share of citations,
31\%.  It receives more astronomy citations per IAU member, per unit
GDP or per head of population than most other countries.  The UK comes
second (11\% of citations), followed by Germany (5\%), Canada (4\%),
Italy (3\%), France (3\%), Japan (2\%), the Netherlands (2\%),
Switzerland (2\%) and Australia (2\%).  The statistics for the other
countries are based on too few papers to permit a meaningful ranking.
The sum for ESO countries (8 members during the period covered) is
17\% (so it should rise to $\sim$ 27\% with the admission to ESO of
the UK). The split 61\% USA, 11\% UK, 18\% continental Europe, 10\%
rest of the world is almost unchanged if just the top 10 most-cited
papers from each year (i.e. 80 papers in all) are considered: 64\%
USA, 14\% UK, 12\% continental Europe, 10\% rest of the world

For a nearly independent measure of impact, we looked at the 452 
observational astronomy/space papers published in Nature 
(i.e. high-impact papers) during 1989-98.
52\% of these had first authors at institutions in the USA,
13\% in the UK,
20\% in continental Europe,
15\% rest of the world, similar to the fractions above.

There were no significant changes in the citation fractions from 
the 10 most-productive countries between 1991-4 and 1995-8.
The statistics are too small to allow a comparison for the remaining
countries.  
For science overall, output is growing fastest in 
Spain, according to Nature (1999), and the fraction of astronomy
papers published from Spain is rising steadily
(S\'{a}nchez \& Benn 2001).

\subsection{Citations per astronomer}

The number of IAU members hosted by a given country is an imperfect
measure of the resources invested in astronomy, because of
country-to-country variations in the way IAU members are put forward.
However, countries with a high citation fraction per IAU member,
e.g. the Netherlands or Switzerland (Fig. 2a), tend also to have high
citation fractions per unit all-science fraction, GDP or population
(Figs. 2b,c,d), suggesting that the scatter about the mean in Fig. 2a
is dominated by differences in mean citation fraction per unit
investment in astronomy, rather than errors in the way that investment
is measured.

Citation fraction per 100 IAU members (column 7 of Table 1) is
relatively high in the USA (2.7\%) and the UK (2.0\%), probably in
part due to the biases mentioned earlier (such as language and
regional biases).  The mean for the other major contributors
(continental Europe, Australia, Canada and Japan) is 0.8 $\pm$ 0.1\%
per 100 IAU members, and varies little from country to country,
despite the large variations in scientific culture, and in
susceptibility to citation bias.  Mean citation counts for different
regions are shown in Table 2.

Within continental western Europe, where the effects of biases against
citation counts are likely to be relatively homogeneous, differences
in citations per astronomer between regions might trace the effects of
different research cultures.  To test this, we split Europe into two
regions, roughly north vs south.  The countries in the north of Europe
(Austria, Belgium, Denmark, Finland, Germany, Netherlands, Sweden and
Switzerland, 1017 IAU members, 7.5 per million of inhabitants)
received a mean of 1.04 $\pm$ 0.10\% of the citations per 100 IAU
members.  The countries of the south, (France, Italy and Spain, 1222
IAU members, 7.9 per head of population) received a mean of 0.56 $\pm$
0.06\% of the citations per 100 IAU members.  The north / south ratio
between these numbers is 1.86 $\pm$ 0.27.  The north / south ratios of
citation fractions per unit GDP and per unit population are 1.52 and
1.76 respectively (with similar fractional errors).  The difference
between north and south is small but mildly significant.  It might be
attributed to a number of factors, e.g.: different fractions of
resources allocated through competitive peer review; relative emphasis
on funding research in research institutions vs universities (see
e.g. May 1997); or openness of competition for posts.  The last of
these is a matter of particular concern in France (e.g. Goodman 2001),
Italy (e.g. Abbott 2001, Chiesa \& Pacifico 2001, Nature 2001, Nature
2002a) and Spain (e.g. Bosch 1998, Bosch 1999, Escartin 1998, Pickin
2001).  For a proxy measure of openness of recruitment, we used the
fraction of university teaching staff who were trained at that
university (Navarro \& Rivero 2001), as measured by Soler (2001) for
zoology and ecology departments.  High `inbreeding' fractions are
likely to reflect a tendency to favour internal candidates, i.e. to
allocate posts on grounds other than scientific merit.  Soler (2001)
measured this fraction for several countries: Spain (88\%), Italy
(78\%), Austria (73\%), France (65\%), Belgium (52\%), Finland (48\%),
the Netherlands (40\%), Denmark (39\%), Sweden (32\%), Switzerland
(23\%), the UK (5\%) and Germany (1\%).  The citation counts per IAU
member are compared in Table 2 with the inbreeding fractions.

\begin{table}
\caption{Citation counts per IAU member, by region}
\begin{tabular}{lrl}
\hline
Countries        &Citations fraction      & Inbreeding  \\
                 &/100 IAU members & fraction (\%)  \\
\hline
US                     & 2.70 $\pm$ 0.10& $-$\\
UK                     & 2.00 $\pm$ 0.20& 5\\
CA                     & 1.90 $\pm$ 0.30& $-$\\
AT BE DK FI            & 1.00 $\pm$ 0.10&22 \\
DE NL SE CH            &              & \\
AU                     & 0.80 $\pm$ 0.20& $-$\\
FR IT ES               & 0.60 $\pm$ 0.10&73 \\
JP                     & 0.50 $\pm$ 0.10& $-$\\
Developing world       & 0.03 $\pm$ 0.01& $-$\\
\hline
\end{tabular}

Column 3 gives for each group of countries the mean 
fraction of teaching staff
who trained at the universities where they currently hold a post
(Soler 2001, see text).

`Developing world' includes
developing countries with more than 100 IAU members in 1998:
Brazil, China (no papers in top 1000), India (no papers in top 1000), 
Russia, Ukraine.
The quoted errors are measurement $\times$ (number of papers)$^{-0.5}$.
\end{table}

The fractions of citations per IAU member in Australia and Japan are similar 
to those in Europe, and are less than those in the USA and UK,
consistent with language bias being less important than
biases arising from being part of a large self-citing community.

The USA and 
Switzerland received the largest fraction of citations per IAU member.
May (1997) found that
Switzerland also leads the world in terms of
numbers of all-science papers or 
citations per head of population.

The citation fraction per IAU member in the developing world
is lower, probably due to many factors, including lack of resources,
loss of expertise abroad, and langauge and other biases mentioned
in the introduction.
In addition, our arbitrary assignment of the citation credit 
for each paper to the country hosting the
first author may bias against developing countries
whose astronomers usually work as part of large international teams.

In Fig. 3 we compare the citation fraction for each country with the
fraction of all astronomy articles published.
The USA dominates, and the ratio of 
the two measures is higher for the USA than for other countries.
When papers from the USA are excluded (Fig. 3b), 
the two fractions for each country are similar,
i.e. number of papers published is a useful proxy for scientific impact.

\subsection{Regional subject preferences}

The sample of 1000 most-cited papers maps regional preferences for
different areas of research.  Of the papers on stars and on our
galaxy, 62\% are published from the USA, 2\% from the UK and 36\% from
other countries.  Similar regional distributions are obtained when
this subset of papers is further sub-divided into papers on hot stars,
on cool stars or on the galaxy.  Papers on external galaxies split
56\% USA, 21\% UK, 23\% other.  Papers on AGN and cosmology split 65\%
USA, 12\% UK, 23\% other.  Within the 1000-paper sample, the USA is
relatively strongest in cosmology, the UK in studies of external
galaxies, Europe in studies of cool stars.

These regional differences are reflected in the way individual
telescopes were used.  During the period studied, 1991-98, the most
productive ground-based telescopes were of 4-m class. 189 of
the 1000 top-cited articles were based on data from such 
telescopes: 143 with first author from the USA, 28 from the UK and 18
from continental Europe. For 4-m telescopes whose users are
predominantly from North America, 98 of the 134 (73\%) papers
are on extragalactic topics.  For 4-m telescopes used by the UK (WHT,
AAT), the fraction is 28 out of 33 (85\%).  For 4-m telescopes whose
users are predominantly from continental Europe (Calar Alto 3.5m, NTT
3.5m and ESO 3.6m), 8 of the 18 papers (44\%) are on extragalactic
topics.

These regional differences are also reflected in the extragalactic
fraction in different journals: 62\% of the subset of the top 1000
papers were published in ApJ, 83\% of those published in MNRAS, 30\%
of those in A\&A.  If galactic and extragalactic astronomers had
different citation practices, these differences could induce a
regional bias in the citation counts.  To test this, we counted the
number of references in 25 of the papers from each of the US
extragalactic and European (non-UK) stellar communities.  The numbers
of references ranged (excluding smallest and largest sixth,
i.e. approximating $\pm$ 1 standard deviation for a gaussian) between
37 and 79 for extragalactic papers and between 24 and 78 for galactic
ones.  This suggests that any difference in citing rate between the
two communities is small, and that the regional differences in
preferred topics won't bias the citation count.

The emphasis on astrophysics, relative to science as a whole, also
varies from country to country.  The ISI web pages listed for each
country the fraction of papers published in each of 22 disciplines,
for the period $\sim$ 1994-8.  The USA published $\sim$ 30\% of all
science and social-sciences papers worldwide, but $\sim$ 40\% of all
astrophysics papers.  In terms of fraction of world output published,
astrophysics ranks 6th in the USA (law is first, with 90\% of world
output, followed by four social-science subjects).  In France,
Germany, Italy, Netherlands, Spain and the UK, astrophysics ranks
first, i.e.  is the discipline in which each country publishes a
larger fraction of world output than in any other discipline.  In
Japan, astrophysics ranks 15th.

\begin{figure}
\centering
\psfig{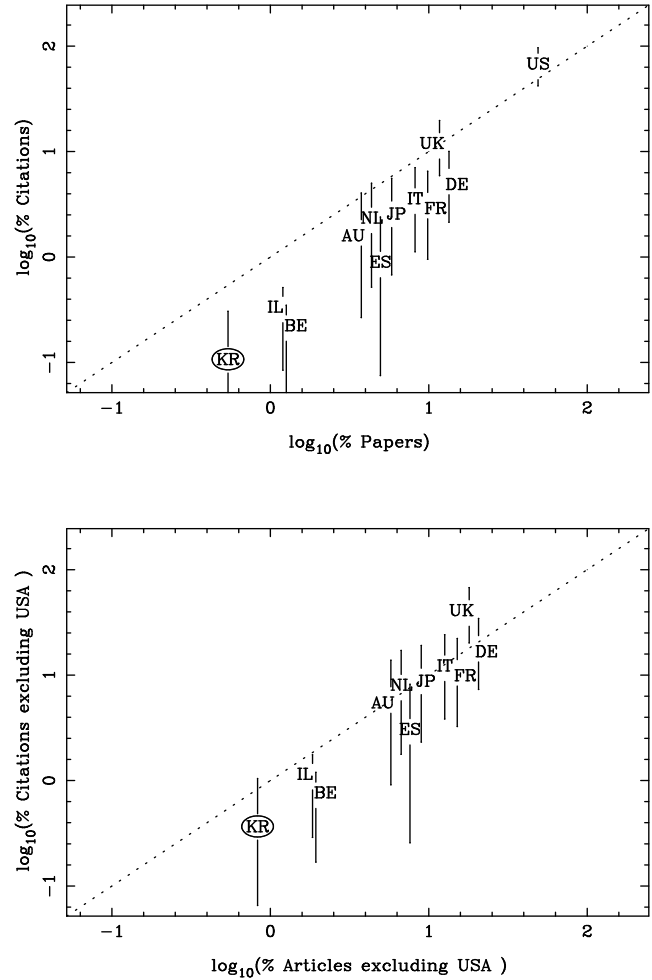}
\caption{
(a) 
Relation between citation fraction 
(to 1000 most-cited articles) and fraction of {\it all} articles
published in astronomy 1991-8 (data from ISI web page).
(b) Same data as (a), but after excluding papers from the USA.
In both panels, the error bars are measurement $\times$
(number of papers)$^{-0.5}$.
}
\end{figure}

\section{Conclusions}

From our analysis of
the 1000 most-cited astronomy papers published
during 1991-98 we conclude:

\begin{enumerate}
\item 61\% of the citations are to papers with first authors at institutions
  in the USA, 11\% in the UK, 5\% in Germany, 4\% in Canada, 3\% in Italy, 3\%
  in France.  17\% are to papers with first authors at institutions in (then)
  ESO countries. Language and other biases favour relative over-citation of
  papers from the large north-American and UK communities.

\item
The number of citations generated by research in developed countries
is proportional to the number of IAU members, over a range of 2 orders
of magnitude, with rms $\approx$ 0.2 dex.  The number of citations per
astronomer is highest for the USA, with 2.7\% of all citations per 100
IAU members.  The number of citations per IAU member is similar for
Canada and the UK.  The mean for continental Europe is 0.8\% per 100
IAU members, and is significantly higher in the north than in the
south.
\item
For most countries, citation fraction is similar to the fraction of
all astronomy papers published by that country, i.e. the latter is a
useful proxy for scientific impact.
\item
Citation fraction is proportional to GDP 
for countries with GDP $>$ $\$$ 15000 per capita.
\item
62\% of the extragalactic papers in the sample are published from the
USA, 15\% from the UK, 23\% from other countries (mainly in Europe).
In stellar astronomy, 62\% of the papers are again published from the
USA, but 2\% from the UK, and 36\% from other countries.  Several
metrics indicate a regional difference in relative emphasis on
extragalactic and stellar research, with extragalactic work dominating
in the USA and UK, and a greater emphasis on stellar work in Europe.

During the period 1991-8 covered by this study, ground-based optical astronomy
was dominated by 4-m class telescopes.  Since then, several 8-m class
telescopes (e.g. Gemini, HET, Keck, Subaru, VLT) have come into use, and more
are on the way (e.g. GTC, SALT).  These will give 8-m access to countries
which previously had only limited access to 4-m telescopes (e.g. Japan) and to
countries whose astronomy communities are still expanding rapidly (e.g.
Spain).  It will therefore be interesting to repeat this study once there are
enough citations to papers from the new 8-m telescopes to permit statistical
analysis.

\end{enumerate}

\begin{acknowledgements}

This research was supported in part by the Euro3D Research Training
Network, founded by the EC, under contract HPRN-CT-2002-00305.
We are grateful to Lutz Wisotzki and Mark McCaughrean for comments. 
\end{acknowledgements}



\end{document}